\documentclass[pra,twocolumn,showpacs,amsmath,amssymb,byrevtex]{revtex4}
\usepackage{amsmath,amssymb,amsthm,graphicx,color}

\newtheorem{Thm}{Theorem}

\newtheorem{Conj}{Conjecture}

\theoremstyle{definition}

\newcommand{\HA}{\mathop{\mathcal{H}}\nolimits}
\newcommand{\T}{\mathop{\mathcal{T}}\nolimits}
\newcommand{\B}{\mathop{\mathcal{B}}\nolimits}

\newcommand{\CA}{\mathop{\mathbb{C}}\nolimits}
\newcommand{\N}{\mathop{\mathbb{N}}\nolimits}
\newcommand{\tr}{\mathop{\mathrm{Tr}}\nolimits}
\newcommand{\I}{\mathop{\mathbb{I}}\nolimits}

\newcommand{\bra}[1]{\langle #1 |}

\newcommand{\ket}[1]{| #1 \rangle}

\newcommand{\ketbra}[2]{| #1 \rangle \langle #2 |}

\begin{document}

\title{How to detect a possible correlation from the information of a sub-system in quantum mechanical systems}
\author{Gen Kimura $^a$}
\email{gen@ims.is.tohoku.ac.jp}
\author{Hiromichi Ohno $^b$}
\email{ohno@math.kyushu-u.ac.jp}
\author{Hiroyuki Hayashi $^c$}
\affiliation{$^a$ Graduate School of Information Sciences, Tohoku University, Aoba-ku, Sendai 980-8579, Japan}
\affiliation{$^b$ Graduate School of Mathematics, Kyushu University, 1-10-6 Hakozaki, Fukuoka 812-8581, Japan}
\affiliation{$^c$ Department of Physics, Waseda University, Tokyo 169-8555, Japan}

\date{\today}
\begin{abstract}
A possibility to detect correlations between two quantum mechanical systems only from the information of a subsystem is investigated. 
For generic cases, we prove that there exist correlations between two quantum systems if the time-derivative of the reduced purity is not zero. 
Therefore, an experimentalist can conclude non-zero correlations between his/her system and some environment if he/she finds the time-derivative of the reduced purity is not zero.  
A quantitative estimation of a time-derivative of the reduced purity with respect to correlations is also given. 
This clarifies the role of correlations in the mechanism of decoherence in open quantum systems.

\end{abstract}

\maketitle

\section{Introduction}
In many contexts in physics, it is important to know the existence (or absence) of correlations \cite{not:cor} of a system of interest $S$ and its environment $E$ (another unknown system). 
For example, in order to achieve a successful quantum information processing, quantum communication or quantum mechanical control, one has to manage system-environment correlations which may enhance the decoherence of the states of the system. 
However, in many cases, we know neither the structure of the environment nor the nature of the interaction of the system. 
Under these circumstances, one has to detect possible correlations between $S$ and $E$, only from the measurements of the system $S$, not from those of the total system $S+E$. 
To do this, if an ensemble of independently identical systems is available, the following well-known criterion \cite{ref:d'Espagnat:BOOK} for quantum systems can be applied:

\noindent{\bf (A) If the system $S$ is in a pure state, then $S$ has no correlations with any other environment $E$.}

\noindent 
From this statement, an experimentalist can safely conclude no correlations with any environment if he/she found his/her (reduced) state in a pure state.  
Indeed, some of the unconditional security proofs of quantum cryptography partially rely on this fact \cite{ref:usp}, where an unknown eavesdropper is assumed to prepare any environment and do anything which is physically allowed. 

Unfortunately, statement (A) is unavailable when the reduced state is in a mixed state.
Indeed, then, any static property of a subsystem cannot provide the information of the correlation, since the same reduced mixed states can be generated from the total states with and without the system-environment correlations \cite{not:propermixedness}. 
Therefore, in such cases, we would need to use dynamical information as well. 
Here what we would like to discuss and try to show is the following statement: 

\noindent{\bf (B) If the time derivative of the purity of $S$ is not zero at time $t=t_0$, there exist non-zero correlations with a certain environment at that time}.  

\noindent If this statement is universally true, this makes an experimentalist possible to confirm non-zero correlations with some environment if he/she found the time derivative of the purity is not zero \cite{not:interact}. 
The purpose of this paper is to investigate statement (B) for arbitrary quantum mechanical systems \cite{not:4} under the usual postulates for (open) quantum mechanics (see, for instance \cite{ref:vN,ref:OQS,ref:Davies}), which include the followings:  

(i) {\bf [State space]} {\it For any quantum mechanical system $S$, there exists a separable Hilbert space $\HA_S$. Any state of $S$ is represented by a density operator $\rho_S$ --- a positive trace class operator on $\HA_S$ with unit trace. }

The purity $P_S$ for $\rho_S$ is defined by
\begin{equation}
P_S = {\rm Tr}_S\{\rho_S^2\}. 
\end{equation}

(ii) {\bf [Composite system]} {\it Let $S$ and $E$ be quantum mechanical systems with Hilbert spaces $\HA_S$ and $\HA_E$. The composite system $S+E$ is associated with the tensor product Hilbert space $\HA_S\otimes \HA_E$. }

For a total density operator $\rho_{tot}$ on $\HA_S \otimes \HA_E$, the reduced states $\rho_S$ and $\rho_E$ for $S$ and $E$ are given by $\rho_S = \tr_E\{\rho_{tot}\}$ and $\rho_E = \tr_S\{\rho_{tot}\} $ where Tr$_S$ and Tr$_E$ are the partial traces with respect to $S$ and $E$, respectively. (In the following, $\rho_S$ and $\rho_E$ always represent the reduced density operators on $S$ and $E$ from the total density operator $\rho_{tot}$.) 
No correlations in a density operator $\rho_{tot}$ on $S+E$ equivalently means that $\rho_{tot}$ is given by a tensor product of the reduced density operators of the two subsystems: 
\begin{equation}\label{eq:prod}
\rho_{tot} = \rho_S \otimes \rho_E. 
\end{equation}

(iii) {\bf [Evolution]} {\it A quantum system $S$ is dynamically isolated or open, and without or with a certain environment $E$, the dynamics of $S$ is eventually described by the von Neumann equation (Schr\"odinger equation) on the total system. Namely, there exists a self-adjoint Hamiltonian $H$ on $\HA_S\otimes\HA_E$ with which the von Neumann equation holds: 
\begin{equation}\label{eq:vNeqNaive}
i \hbar \frac{d}{dt} \rho_{tot}(t) = [H,\rho_{tot}(t)], 
\end{equation}
where $\rho_{tot}(t)$ is a density operator on $\HA_S\otimes\HA_E$ at time $t$. (In the following, we set Planck's constant $\hbar$ to be $1$.)}

Notice, however, that there appears a domain-problem when $H$ is an unbounded operator \cite{ref:RS}. 
To avoid the problem, it is generally adopted in the axiomatic approach of quantum mechanics that the dynamics is governed by a unitary time evolution:   
\begin{equation}\label{eq:vNgeneral}
\rho_{tot}(t) = U_t \rho_{tot} U^\dagger_t, 
\end{equation}
where $\rho_{tot}$ is an initial density operator at $t=0$ and $U_t$ is a unitary operator given by $U_t = e^{-i H t}$ (for a time-independent Hamiltonian $H$). Then, for any density operator $\rho_{tot}$, the dynamics \eqref{eq:vNgeneral} is applied without any problem such as a domain-problem. 
In this paper, we assume a unitary dynamics \eqref{eq:vNgeneral} for an isolated quantum system.

In a formal analysis, statement (B) for quantum mechanical systems can be proved in the following way: 
Let the time-derivative of the purity of a quantum system $S$ at $t=t_0$ is not zero. 
Since the purity does not change in an isolated system, $S$ should be an open system interacting with some environment $E$.  
Let $H$ be a self-adjoint Hamiltonian on $\HA_S\otimes\HA_E$ which reads the von Neumann equation \eqref{eq:vNeqNaive}. 
Assume that there are no correlations at $t=t_0$, namely the initial density operator takes a product form $\rho_{tot} = \rho_S \otimes \rho_E $. Then, from the von Neumann equation, we observe,
\begin{eqnarray}\label{eq:proof1}
P^\prime_S(t_0) &\equiv & \frac{d}{dt} P_S(t)\Big|_{t=t_0}  \nonumber \\ 
&=& 2 \tr_S\left\{\rho_S(t) \frac{d}{dt}\rho_S(t)\Big|_{t=t_0}\right\} \nonumber \\ 
&=& -2 i \tr_S\left\{ \rho_S \tr_E [H,\rho_S\otimes\rho_E] \right\}\nonumber \\  
&=& -2 i \tr_{SE} \left\{\rho_S \otimes \I_E [H,\rho_S \otimes \rho_E] \right\}\nonumber \\ 
&=& 0,   
\end{eqnarray}
where the cyclic property \cite{ref:trace} of the trace $\tr_{SE}$ has been used to estimate the last equality. Therefore, by contradiction, we conclude that $\rho_{tot}$ has non-zero correlations at $t=t_0$. (In the following, the notation of the Newton's difference quotient such as $P^\prime_S(t_0) \equiv \frac{d}{dt} P_S(t)\Big|_{t=t_0}$ will be used.) 
It is worthy to notice that, although use has been made of a Hamiltonian in the proof, experimentalists do not have to know anything about environments including the way how they are interacting with their systems. 
Instead, only thing they have to believe is the postulates of quantum mechanics such as postulates (i),(ii), and (iii).   
 
Estimation \eqref{eq:proof1}, however, is still rough without sufficient mathematical rigor, especially for the case of infinite dimensional Hilbert spaces. 
Moreover, if the Hamiltonian is described by an unbounded operator, we have to deal with the domain carefully, which makes the statement quite non-trivial. 
In the following, we discuss the validity of statement (B) including infinite dimensional Hilbert spaces in a careful manner.   
In Sec.~\ref{sec:b}, we provide a rigorous version of statement (B) and show more general statement (Theorem \ref{thm:strong}) in the case of bounded Hamiltonians, which quantitatively generalize statement (B). 
This shows how purity changes under the existence of correlations, and 
hence clarifies the role of correlations in the mechanism of decoherence in open quantum systems. 
In Sec.~\ref{sec:ub}, we discuss statement (B) in the case of unbounded Hamiltonians and show a certain counter example. 
Finally, we slightly modify the statement (B) to be correct (Theorem \ref{thm:weaker}) for the case of unbounded Hamiltonians. 
This is done by adding an assumption of a finite variance of a total energy, and hence we conclude that statement (B) is universally valid for all the generic cases. 
Sec.~\ref{sec:CD} closes the paper with some concluding remarks and discussion.   
 
\section{The Case of Bounded Hamiltonians --- Quantitative Estimation of Statement (B)}\label{sec:b}

In this section, we discuss statement (B) including infinite dimensional cases with mathematical rigor, but for the case of bounded Hamiltonians. We obtain a useful theorem which generalizes statement (B) in a quantitative manner (Theorem \ref{thm:strong}). 
As usual when discussing open quantum systems \cite{ref:OQS}, we shall divide a total Hamiltonian $H$ into the sum of free Hamiltonians $H_S$ and $H_E$ for systems $S$ and $E$ and an interaction Hamiltonian $H_{int}$: 
\begin{equation}\label{eq:H}
H = H_S\otimes \I_E + H_{int} + \I_S \otimes H_E.  
\end{equation}
We assume $H_S$, $H_E$ and $H_{int}$ are bounded self-adjoint operators on $\HA_S$, $\HA_E$, and $\HA_S\otimes\HA_E$, respectively, and hence $H$ is also a bounded self-adjoint operator on $\HA_S\otimes \HA_E$. 

In order to quantify correlations between $S$ and $E$ in a state $\rho_{tot}$, we use quantum mutual information \cite{ref:Lindblad,not:op}:
$$
I(\rho_{tot}) \equiv \tr_{SE}\left\{\rho_{tot}\log \rho_{tot} - \rho_{tot}\log \rho_S\otimes\rho_E\right\},
$$
where $\rho_S$ and $\rho_E$ are reduced density operators on $S$ and $E$, respectively. Notice that $I(\rho_{tot}) \ge 0$, and $I(\rho_{tot}) = 0$ iff $\rho_{tot}$ has no correlations. Notice also that \cite{ref:PO}
\begin{equation}\label{eq:ineq}
||\rho_{tot} - \rho_S\otimes\rho_E||_1 \le 2 I(\rho_{tot}),
\end{equation}
where $||\cdot||_1$ is the trace norm $||W||_1 \equiv \tr_{SE}\left\{\sqrt{W^\dagger W}\right\}$ \cite{ref:trace}.  

For any density operator $\rho_{tot}$ on $\HA_S\otimes\HA_E$, we define the correlation operator $\rho_{cor}$ \cite{ref:HKO} by 
\begin{equation}\label{eq:corop}
\rho_{cor} \equiv \rho_{tot} - \rho_S \otimes \rho_E,
\end{equation}
which is a trace class operator on $\HA_S\otimes\HA_E$. By definition, it holds that $\rho_{cor}=0$ iff $\rho_{tot}$ has no correlations with a product form \eqref{eq:prod}. Since $\tr_E \{\rho_S \otimes \rho_E\} = \rho_S$, it follows
\begin{eqnarray}\label{eq:corop2}
\tr_E \{\rho_{cor}\} = 0.   
\end{eqnarray}   
We have the following quantitative estimation of a time-derivative of the reduced purity: 
\begin{Thm}\label{thm:strong} 
Let $S$ and $E$ be quantum mechanical systems where the total system $S+E$ is a closed system. 
Let $H$ be a total Hamiltonian and $\rho_{tot}$ be a density operator at $t=t_0$. 
If $H$ is bounded with the form \eqref{eq:H}, then the reduced purity $P_S(t)$ is time-differentiable at $t=t_0$ and 
\begin{equation}\label{eq:ComIntCor}
P^\prime_S(t_0) = -2i \tr_{SE}\left\{\rho_S\otimes \I_E [H_{int},\rho_{cor}]\right\}.
\end{equation}
The absolute value of the time-derivative is bounded from above by 
\begin{subequations}\label{eq:strong} 
\begin{eqnarray}
|P^\prime_S(t_0)| &\le& 2||\rho_S|| \ ||[H_{int},\rho_{cor}]||_1,  \label{eq:stronga}  \\
                &\le& 4 ||H_{int}|| \ ||\rho_{cor}||_1, \label{d} \\
                &\le& 8 ||H_{int}|| I(\rho_{tot}), \label{i}
\end{eqnarray}
\end{subequations}
where $||\cdot|| $ denotes the operator norm \cite{ref:trace}.
\end{Thm}
{\it Proof}. \ 
Notice that $[H,\rho_{tot}(t)]$ is a trace class operator due to an ideal property of trace class operators \cite{not:formula} and the von Neumann equation \eqref{eq:vNeqNaive} holds \cite{ref:Davies} for any density operator where the time derivative is defined with respect to the trace norm.  
Therefore, by observing the inequalities \cite{ref:trace}:  
\begin{eqnarray}\label{eq:formula}
|\tr\{A \rho\}| &\le& ||A \rho||_1 \le ||A|| \ ||\rho||_1, \nonumber \\
 && \ (\forall A \in \B(\HA), \ \rho \in \T(\HA)), 
\end{eqnarray} 
and $||\rho_S(t)\otimes \I_E|| \le 1$ \cite{not:opnorm}, $P_S(t)$ is differentiable for any time $t$ and we have  
$$
P^\prime_S(0) = -2i \tr_{SE}\left\{\rho_S\otimes \I_E [H,\rho_{tot}]\right\}. 
$$
By the cyclic property of the trace \cite{not:cyclic}, it follows $\tr_{SE}\left\{\rho_S\otimes \I_E [H,\rho_{S}\otimes\rho_E]\right\} = \tr_{SE}\left\{[\rho_{S}\otimes\rho_E,\rho_{S}\otimes\I_E] H\right\} = 0$, and therefore, we have
$$
P^\prime_S(0) = -2i \tr_{SE}\left\{\rho_S\otimes \I_E [H,\rho_{cor}]\right\}. 
$$
Moreover, since $\tr_{SE} \left\{ \rho_S\otimes \I_E [H_S\otimes \I_E,\rho_{cor}]\right\} = \tr_S\{\rho_S[H_S, \tr_E \rho_{cor}]\} = 0$ from \eqref{eq:corop2}, and $\tr_{SE} \{\rho_S\otimes \I_E [\I_S\otimes H_E,\rho_{cor}]\} = \tr_{SE} \{[\rho_S\otimes \I_E,\I_S\otimes H_E]\rho_{cor}\} = 0 $ again by the cyclic property of the trace, we obtain \eqref{eq:ComIntCor}. 
From \eqref{eq:formula}, $[H_{int},\rho_{tot}] \in \T(\HA_S\otimes\HA_E)$ and $||\rho_S\otimes \I_E|| = ||\rho_S||$, we have
$$
|P^\prime_S(0)| \le 2 ||\rho_S|| \ ||[H_{int},\rho_{cor}]||_1. 
$$  
The second inequality \eqref{d} follows from the triangle inequality for the trace norm, $||\rho_S|| \le 1$, and again \eqref{eq:formula}.  The third inequality \eqref{i} follows from \eqref{eq:ineq}.  

\hfill{{\it QED}}

\noindent  Theorem \ref{thm:strong} provides a quantitative estimation of a time-derivative of the reduced purity in terms of the amount of correlations $I(\rho_{tot})$ and the strength of interaction $||H_{int}||$ \cite{ref:Miyadera}. It is worth to notice that the inequalities \eqref{eq:strong}s include the following well-known fact \cite{not:interact}: the purity of system does not change without an interaction with an environment. 
Indeed, experimentalists usually confirm the existence of an interaction between the system and some environment, if they find the reduced purity not to be constant. However, not only that, \eqref{eq:strong}s imply that correlations play an essential role in changing the purity even in the existence of an interaction. Moreover, Eq.~\eqref{eq:ComIntCor} implies that the commutator between the interaction Hamiltonian and the correlation operator is essential for the changes of purity, or decoherence.  
 
From Theorem \ref{thm:strong}, we obtain a rigorous version of statement (B): 
\begin{Thm}\label{thm:weak}
With the same assumptions as in Theorem \ref{thm:strong}, if there are no correlations at $t=t_0$: $\rho_{tot} = \rho_S \otimes \rho_E$, then $P_S(t)$ is time-differentiable at $t=t_0$ and $P^\prime_S(t_0) = 0$. 
In other words, if the time-derivative of the reduced purity is not zero, then there exists a non-zero correlation between $S$ and $E$ at that time.  
\end{Thm} 
{\it Proof}. \ Since $\rho_{tot} = \rho_S \otimes \rho_E$ implies $\rho_{cor} = 0$, we have $P^\prime_S(t_0) = 0$ from inequality \eqref{eq:stronga}. 
\hfill{{\it QED}}

It should be noticed that the opposite statement does not generally true. (For instance, if $H_{int}=0$, we have $P^\prime_S(t_0) = 0$ even in the presence of correlations.) 
Therefore, it is incorrect to infer no correlations when the time-derivative of the reduced purity is zero. 
Notice also that the above theorems do not contradict with the results in Ref.~\cite{ref:1} where we have shown that an effect of an initial correlation does not appear in van Hove's limit (the weak coupling limit) and therefore system $S$ behaves as if the total system started from the factorized initial state. 
Indeed, this is true only for the van Hove time scale $\tau = \lambda^2 t$ where $\lambda \ll 1$ is a coupling constant, and in much shorter time scales than $\tau$, we can find a difference between no correlations and non-zero correlations as we have seen in the above Theorems. (See also \cite{ref:HKO} for an effect of an initial correlation.)


\section{The Case of Unbounded Hamiltonians --- Counter Examples of Statement (B)}\label{sec:ub}
 
In the previous section, we have confirmed that statement (B) is universally true for any bounded Hamiltonian. 
However, Hamiltonians are generally unbounded, especially from above, like that of the harmonic oscillator. 
Notice that, although the quantitative estimation \eqref{eq:strong} in Theorem \ref{thm:strong} turns out to be trivial when $||H_{int}|| = \infty$, we may still expect the validity of Theorem \ref{thm:weak}, i.e., statement (B). 
In this section, we discuss statement (B) in the case of unbounded Hamiltonians.  
However, as we shall see below, the statement itself can be generally broken down. In the following, we provide a counter example of statement (B).  

\vspace{3mm}
\noindent[{\it Counter Example of statement (B)}]
\vspace{3mm}

Let our system be described by $\HA_S = \HA_{S_1}\otimes \HA_{S_2}$ where $\HA_{S_1}$ is a separable Hilbert space with an infinite dimension, and $\HA_{S_2}$ is a $2$ dimensional Hilbert space, $\HA_{S_2} \simeq \CA^2$. (For instance, it is a system of a non-relativistic electron with spin $1/2$.)
To avoid a technical complexity, we use the simplest environment $\HA_E \simeq \CA^2$, which is also a $2$ dimensional Hilbert space. 
  
Assume that initially the total system is in a state $\rho_{tot} = \rho_S \otimes \rho_E$ which has no correlations, where 
\begin{equation}\label{eq:inist}
 \rho_S = \sum_{n=1}^\infty p_n \ketbra{\phi_n}{\phi_n} \otimes \ketbra{s_1}{s_1}, \ \rho_E = \ketbra{e_1}{e_1},
\end{equation}
with $p_n \ge 0, \ \sum_{n=1}^\infty p_n = 1$, and orthonormal bases $\{\ket{\phi_n}\}_{n=1}^\infty$, $\{\ket{s_n}\}_{n=1}^2$, and $\{\ket{e_n}\}_{n=1}^2$ of $\HA_{S_1}$, $\HA_{S_2}$, and $\HA_E$, respectively. 

We use the following Hamiltonian $H$, whose spectral decomposition reads  
$$
H = \sum_{n=1}^\infty \sum_{k=1}^4 h_{n k } \ketbra{\phi_n \otimes \chi_k}{\phi_n \otimes \chi_k}, 
$$
with eigenvalues (point spectra) $h_{n1}=0, h_{n2}=h_{n3} = h_n, h_{n4} = 2 h_n $ with $h_n \ge 0 \ (n \in \N)$, where $\{\ket{\chi_k}\}_{k=1}^4$ is an orthonormal basis of $\HA_{S_2}\otimes\HA_E$ given by 
\begin{eqnarray*}
\ket{\chi_1} &\equiv& \frac{1}{\sqrt{2}} (\ket{s_1\otimes e_1} + i \ket{s_2 \otimes e_2}), \\
\ket{\chi_2} &\equiv& \ket{s_2 \otimes e_1}, \ \ket{\chi_3} \equiv \ket{s_1 \otimes e_2}, \\
\ket{\chi_4} &\equiv& \frac{1}{\sqrt{2}} (\ket{s_1 \otimes e_1} - i \ket{s_2 \otimes e_2}). 
\end{eqnarray*}
By the above spectral decomposition, it is easy to see that $H$ is a positive self-adjoint operator on $\HA_S\otimes\HA_E$, which is unbounded when the sequence $\{h_n\}$ is not bounded from above. 
The time evolution map $U_t = \exp(- i H t)$ is given by 
$$ 
U_t = \sum_{n=1}^\infty \ketbra{\phi_n}{\phi_n}\otimes X^n_t, 
$$
where $X^n_t \equiv \ketbra{\chi_1}{\chi_1} + e^{-i h_n t} (\ketbra{\chi_2}{\chi_2} + \ketbra{\chi_3}{\chi_3}) + e^{-i 2 h_n t} \ketbra{\chi_4}{\chi_4}$.  
By \eqref{eq:inist} we have   
\begin{eqnarray}
\rho_{tot}(t)&=&  U_t \rho_{tot} U_t^\dagger \nonumber \\
 &=& \sum_{n=1}^\infty p_n \ketbra{\phi_n}{\phi_n} \otimes \ket{X^n_t s_1\otimes e_1}\bra{X^n_t s_1\otimes e_1}\nonumber,  
\end{eqnarray}
where $\ket{X^n_t s_1\otimes e_1} = e^{-i h_n t} (\cos (h_n t) \ket{s_1 \otimes e_1} - \sin (h_n t) \ket{s_2 \otimes e_2 })$. By taking a partial trace over $E$, we have $ \rho_S(t) =  \sum_{n=1}^\infty p_n \ketbra{\phi_n}{\phi_n} \times ( \cos^2 (h_n t) \ketbra{s_1}{s_1} + \sin^2 (h_n t) \ketbra{s_2}{s_2})$. 
From this, we obtain an analytical form of the reduced purity:   
\begin{eqnarray}
P_S(t) = \sum_{n=1}^\infty p_n^2 (\cos^4 (h_n t)+ \sin^4 (h_n t)) \nonumber \\
 = P_S(0) - \frac{1}{2} \sum_{n=1}^\infty (p_n \sin(2 h_n t))^2 \nonumber \\
 = \frac{3}{4}P_S(0) + \frac{1}{4}\sum_{n=1}^\infty p_n^2 \cos[4 h_n t], \label{eq:OhnoPurity}  
\end{eqnarray} 
where $P_S(0) = \sum_{n=1}^\infty p_n^2$. 
\begin{figure}[t]
\includegraphics[height=0.55\textwidth]{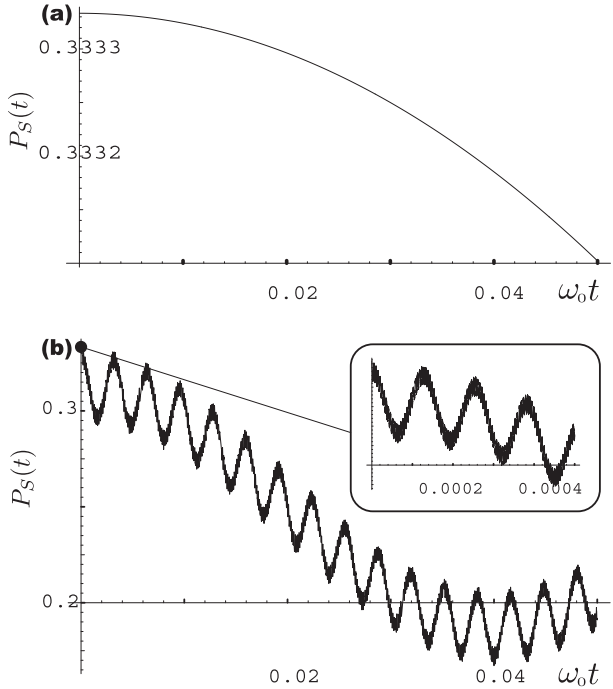}
\caption{Time evolution of the reduced purity \eqref{eq:OhnoPurity} for (a) $p_n = \frac{1}{2^n}, h_n = \frac{n}{4}$ and (b) $p_n = \frac{1}{2^n}, h_n = \frac{25^n \pi}{4}$, with a unit of time $\omega_0 \equiv \hbar/E_0$.  
Notice that in both cases the Hamiltonians are unbounded from above. 
One sees the flat time derivative at $t=0$ in (a) which makes statement (B) to be true, while one sees non-differentiability in (b) which breaks down statement (B). }\label{fig:wei4}
\end{figure}
Therefore, if the infinite sum in \eqref{eq:OhnoPurity} and the time-derivative is commutative, we obtain $P^\prime_S(0) = 0$ and statement (B) holds. 
For instance, let $p_n = \frac{1}{2^n}$, and $h_n = \frac{n E_0}{4}$ with a unit of energy $E_0$. Then, since $|\frac{d}{dt} p_n^2 \cos[4 h_n t]| = |\frac{n E_0 \sin[nE_0 t]}{4^n}| \le \frac{nE_0}{4^n}$ and $\sum_{n=1}^\infty \frac{nE_0}{4^n} < \infty$, it follows that $\sum_{n=1}^\infty p_n^2 \cos[4 h_n t]$ is differentiable with respect to $t$ and we have $\frac{d}{dt} \sum_{n=1}^\infty p_n^2 \cos[4 h_n t] = \sum_{n=1}^\infty 4 p_n^2 h_n \sin[4 h_n t]$. 
Hence, this example satisfies statement (B) even in the case of unbounded Hamiltonians. See FIG.~\ref{fig:wei4} (a). (In the following, we set $E_0 $ to be $1$. )

However, we can construct a counter example of statement (B) in the sense that $P_S(t)$ is not differentiable with respect to $t$ at $t=0$ even when an initial state is given in a product form. We provide an interesting example that $P_S(t)$ is continuous but not differentiable at anytime $t$ by connecting the reduced purity to the so-called Weierstrass function $f(t;a,b)$ \cite{ref:Wei}, defined by 
$$
f(t;a,b) = \sum_{n=0}^\infty a^n \cos (b^n \pi t),  
$$
with two parameters $0 < a <1$ and positive odd integer $b$ satisfying $ab > 1 + \frac{3}{2}\pi$. It is known that the function is continuous everywhere but differentiable nowhere with respect to $t$. From the form of \eqref{eq:OhnoPurity}, a proper choice of $p_n$ and $h_n$, for instance, $p_n = \frac{1}{2^n}$ $h_n = \frac{25^n }{4}\pi$, makes $P_S(t)$ an essentially Weierstrass function: 
\begin{equation}\label{eq:purityWei}
P_S(t) = \frac{1}{4}( 1 - \cos (\pi t) + f(t;\frac{1}{4},25)),  
\end{equation}
(See FIG.~\ref{fig:wei4} (b).) This provides a counter example of statement (B). Namely, even with a product initial state, a time derivative of the purity is not necessarily zero; though this case just provides a case of a non-existence of the time-derivative.  

Therefore, in the case of unbounded Hamiltonians, we need to modify our statement (B). Indeed, the following weaker statement can be proved to be true: 
\begin{Thm}\label{thm:weaker} Let $H$ be a self-adjoint Hamiltonian bounded from below, but not necessarily bounded from above. Let $\rho_{tot}$ be a density operator at $t=t_0$. {\bf If the variance of $H$ with respect to $\rho_{tot}$ is finite}, then 
$$ 
\rho_{tot} = \rho_S \otimes \rho_B \ \Rightarrow 
P^\prime_S(t_0)  = 0.  
$$
\end{Thm} 
\noindent The assumption of the boundedness of the Hamiltonian from below is physically required so that the system to be stable. Hence, even when the Hamiltonian $H$ is unbounded, statement (B) is correct provided that the total state has a finite variance of $H$. In fact, it is easy to see that the variance of $H$ is infinite for the initial state used for the counter example in \eqref{eq:purityWei}. 

To avoid redundant technical difficulties when dealing with unbounded Hamiltonians, in the present paper, we do not give a proof of Theorem \ref{thm:weaker}. Instead, we just notice the followings: First, a finiteness of the variance of $H$ with respect to a pure state $\rho_{tot} = \ketbra{\psi}{\psi}$ is equivalent to that $\ket{\psi}$ is in the domain of $H$. Therefore, from the mathematical point of view, the assumption of a finiteness of the variance of $H$ allows us to avoid a domain-problem for unbounded operators. Second, the von Neumann equation holds when the variance of $H$ is finite, which is the essential reason for the Theorem \ref{thm:weaker} to be correct \cite{not:ce}. We plan to discuss and provide a systematic investigation for the case of unbounded Hamiltonians in the forthcoming paper, including a complete proof of Theorem \ref{thm:weaker}. 

\bigskip 

\section{Concluding Remarks and Discussion}\label{sec:CD}
 
We have discussed the problem how one can detect possible correlations between the system of interest $S$ and an environment from the knowledge (by observations) of the system $S$ only. 
We conjectured statement (B), from which one can conclude non-zero correlations with some environment when the time derivative of the reduced purity is not zero. 
In some sense, it is a counterpart of statement (A); one can conclude no correlations when the reduced purity is $1$ using statement (A), while one can conclude correlations when the time derivative of the reduced purity is not zero. 
For instance, an experimentalist first can use statement (A), and if his/her state is in a pure state, he/she can conclude no correlations. 
If the state is in a mixed state, then he/she can use statement (B) and check the time-derivative of the purity. 
If the time-derivative is not zero, he/she can conclude the existence of correlations, provided that statement (B) is universality true. 
In this paper, we have investigated the validity of statement (B) for arbitrary quantum mechanical systems. 
When the total Hamiltonian is bounded, we proved it to be universally correct (Theorem \ref{thm:weak}), by giving a more general statement (Theorem \ref{thm:strong}) which quantitatively implies statement (B). 
Theorem \ref{thm:strong} also clarifies the cause of a purity-change (decoherence) due to an interaction and correlations.   
However, when the total Hamiltonian is unbounded, we have also shown a counter example of statement (B). 
In the example, the reduced purity evolves essentially as a Weierstrass function even with a product initial state, whence the differentiability of the reduced purity has been broken down in statement (B). 
Therefore, a certain modification is necessary for statement (B). 
If one considers a state with a finite variance of energy as a natural realization in nature, one can conclude the universality of statement (B) for all the generic states in that sense.  
However, considering our original goal to estimate a possible correlation, especially for the situation where we do not know anything about environment (other than our theoretical knowledge of quantum theory), it is preferable to assume nothing additional for an environment \cite{not:QF}. 
In order for this, another plausible conjecture will be   
\begin{Conj}
$$
\exists \ P^\prime_S(0) \ and \ P^\prime_S(0) \neq 0 \Rightarrow \rho_{tot} \neq \rho_S \otimes \rho_B.
$$
\end{Conj} 
\noindent If this is correct, it turns out that one can conclude non-zero correlations if one finds non-zero time derivative (including the differentiability) of the reduced purity. 
In this direction, in the forthcoming paper, we will discuss statement (B) including a complete proof of Theorem \ref{thm:weaker} and an investigation of the above conjecture. Also the case of a quantum field by using an algebraic formalism of quantum fields \cite{ref:QFalg} will be presented elsewhere.  

{\bf Acknowledgement}

We are grateful to Profs. S. Pascazio, I. Ohba, S. Tasaki, H. Nakazato, M. Ozawa, and F. Hiai for their continued encouragements and helpful advices. 
We would like to thank Drs. M. Mosonyi, M. Hotta, K. Yuasa, K. Imafuku, and P. Facchi for their fruitful comments and useful discussions. In particular, we appreciate Profs. Pascazio, Tasaki, Nakazato, and Dr. Yuasa for their careful readings of the manuscript prior publication and Dr. Mosonyi for the useful discussion about the validity of the von Neumann equation. 
This research is supported by the Grant-in-Aid for JSPS Research Fellows.


\begin{thebibliography}{00}
\bibitem{not:cor}
In this paper, we mean a correlation as the statistical correlation in a state of composite physical system $S+E$:
We say there are no correlations between $S+E$ (a statistical independence) iff the joint probability distribution between any two observables $O_S$ from $S$ and $O_E$ from $E$ is a product of the probability distributions for $O_S$ and $O_E$. Otherwise, we say that there exist non-zero correlations.
\bibitem{not:propermixedness}
Some researchers might take a stance that all the mixedness originates from correlations with some environment, 
and eventually the total system should be always described by a pure state. 
If this is universally true, then statement (A) is enough to conclude the existence of correlations when system $S$ is in a mixed state. 
However, our capacity is more general and statement (B) is still useful even if there exists mixedness not originated from correlations.
\bibitem{not:interact}
Notice that an experimentalist would usually conclude the existence of non-zero interaction with some environment when the time derivative of the purity of $S$ is not zero, since the purity of system $S$ does not change if $S$ is isolated. 
Compared to this, statement (B) allows experimentalists to confirm non-zero correlations with some environment. (See Theorem \ref{thm:strong} below.)
\bibitem{not:4}
In a rigorous sense, in this paper, we only treat an environment which is a quantum mechanical system with a fixed separable Hilbert space. Therefore, quantum fields with infinite degrees of freedom are not included. These cases will be investigated in the forth coming paper.
\bibitem{not:op}
In finite dimensional cases, it has been shown that this quantity has an operational meaning as the amount of work to erase both the quantum and classical correlations \cite{ref:GPW}.
\bibitem{not:formula}
Let $\T(\HA)$ be the set of all the trace class operators on a Hilbert space $\HA$. Then, $\T(\HA)$ has an ideal property in $\B(\HA)$, the set of all the bounded operators on $\HA$: Namely, it holds that $A \rho, \rho A \in \T(\HA) $ for any $A \in \B(\HA),\rho \in \T(\HA)$ \cite{ref:trace}.
\bibitem{not:opnorm}
Note that $||\rho|| \le ||\rho||_1$ for any $\rho \in \T(\HA)$ \cite{ref:trace}. Since $\T(\HA_S) \ni \rho_{S}(t) \ge 0$ and $||\I_E|| =1$, we have $||\rho_{S}\otimes \I_E|| = ||\rho_{S}|| \ ||\I_E|| \le ||\rho_{S}||_1 = \tr \{ \rho_S \} =1 $.
\bibitem{not:cyclic}
For any $A \in \B(\HA), T \in \T(\HA)$, we have $\tr\{ A T\} = \tr\{ T A \}$ \cite{ref:trace}.
\bibitem{not:ce}
Indeed, the indifferentiability of \eqref{eq:purityWei} is essentially due to the fact that the von Neumann equation does not hold in this example. However, it is worth noticing that we can provide an example which satisfies statement (B) even when von Neumann equation does not hold.
\bibitem{not:QF}
Moreover, in order to include the case where an environment is a quantum field which has infinite degrees of freedom, the assumption of the environment to be described by a particular fixed Hilbert space should also be removed.
\end{thebibliography}
\end{document}